# Towards a function-scalable quantum network with multiplexed energy-time entanglement

Xiao Xiang[1,3,6], Jingyuan Liu[2,6], Bingke Shi[1,3], Huibo Hong[1,3], Xizi Sun[4], Yuting Liu[1,3], Runai Quan[1,3], Tao Liu[1,3,5], Shougang Zhang[1,3,5], Wei Zhang[2,4,†], Ruifang Dong[1,3,5,‡]

[1]Key Laboratory of Time Reference and Applications, National Time Service Center, Chinese Academy of Sciences, Xi'an 710600, China
[2]State Key Laboratory of Low-Dimensional Quantum Physics, Frontier Science Center for Quantum Information, Electronic Engineering Department, Tsinghua University, Beijing 100084, China
[3]School of Astronomy and Space Science, University of Chinese Academy of Sciences, Beijing 100049, China
[4]Beijing Academy of Quantum Information Sciences, Beijing 100193, China
[5]Hefei National Laboratory, Hefei 230088, China
[6]These authors contributed equally: Xiao Xiang, Jingyuan Liu
†zwei@tsinghua.edu.cn
‡dongruifang@ntsc.ac.cn

## Abstract

Quantum networks, which hinge on the principles of quantum mechanics, are revolutionizing the domain of information technology. The vision for quantum networks involves the efficient distribution and utilization of quantum resources across a network to support a variety of quantum applications. However, current quantum protocols often develop independently, leading to incompatibilities that limit the functional scalability of the network. In this paper, we showcase a compatible and complementary implementation of two distinct quantum applications, quantum time synchronization and quantum cryptography, by multiplexing the same energy-time entangled biphotons and quantum channel. A proof-of-principle experiment between two independent nodes across a 120 km fiber-optic link is demonstrated, which achieve sub-picosecond synchronization stability based on the quantum two-way time transfer protocol. Simultaneously, this synchronization provides the required timing for implementing dispersive-optic quantum key distribution with an average finite-size secure key rate of 73.8 bits per second, which can be employed to safeguard the security of the transferred timing data. Furthermore, thanks to the compatibility, potential asymmetric delay attacks in the link, which are detrimental to the accomplishment of secure key distribution, can be effectively mitigated by the parallel quantum time synchronization procedure. Our demonstration marks a substantial leap towards unlocking the full potential of energy-time entanglement and paves the way for a resource-efficient, function-scalable, and highly compatible quantum network.

## Introduction

The concept of quantum network[1–3], or the more comprehensive quantum internet, heralds a revolutionary transformation in information technology. It capitalizes on the distinct characteristics of quantum mechanics to reshape global communication, computing, and sensing. This novel domain transcends the limitations of classical networks, endowing itself with unparalleled security, computational capabilities and precision. As a cornerstone in quantum networks, entanglement distribution holds the ability of establishing correlated quantum states among distant nodes, which are attracting extensive attention and developments on long-haul fiber and free-space links.[4–8] Given these advancements, entanglement-based quantum network is regarded as a promising platform for supporting advanced quantum protocols beyond the trusted repeater stage[3,8,9]. Within quantum networking protocols, high-precision time synchronization is vital for aligning the arrival times of entangled photon events at each quantum node. To transfer the timing signal, either a separate fiber channel or the coexistence in the same fibers is usually adopted.[10,11] The former requires extra fiber resource consumption, while the latter suffers from the issue of contaminating the entangled photons by classical signals. To address these challenges and meet the growing demands of scalable quantum networks, the need for more compatible time synchronization methods has become increasingly apparent. Additionally, the isolated development of quantum protocols often leads to interface incompatibilities and inefficient use of quantum resources.[12]

Among various types of entanglement, energy-time entanglement stands out due to its high versatility in fields across quantum communication, computation, and metrology. For example, by exploiting the strong temporal correlation of energy-time entangled biphotons, the quantum two-way time transfer (Q-TWTT)[13] protocol has been successfully demonstrated in both metropolitan fiber and free-space links, showing the ability of achieving time synchronization stability far below 1 picosecond[14,15]. These results further solidate its promising prospective of providing high-precision and secure timing for global quantum networking[16] and scenarios in global positioning system denied environments[17]. Leveraging energy-time entangled photon pairs, the dispersive-optic QKD (DO-QKD) protocol[18–20] offers the advantage of source-independent security and the potential for device-independent security. It also provides a technique for constructing large-scale quantum networks with fully-connected topology. Furthermore, energy-time entanglement has been used as a critical component in the development of quantum logic gates[21], offering a robust platform for quantum state manipulation, which are essential for the advancement of quantum computing technologies. Benefitting from its unprecedented versatility, energy-time entanglement is extremely expected to serve as a unified quantum resource, enabling the concurrent implementation of various quantum information technologies within a single network.

In this work, we present a proof-of-principle demonstration for the compatible and complementary realization of quantum time synchronization and quantum cryptography with multiplexed energy-time entanglement over a 120 km fiber-optic link. Employing the Q-TWTT protocol, quantum time synchronization has been successfully implemented to align two independent quantum nodes with sub-picosecond time stability. The synchronization achieved supplies the requisite timing for the DO-QKD accomplishment. A finite-size secure key rate of (73.8±15.7) bits per second was achieved over a 10-hour measurement period, which then can be used to enhance the data layer security of the quantum time synchronization system. Furthermore, facilitated by the dynamic phase compensation within the quantum time synchronization procedure, the DO-QKD acquires the robustness against asymmetric delay attacks. Under malicious asymmetric delay attacks ranging from 10 ps to 120 ps, the normalized secure key rate remained around 80%. These findings represent a significant leap towards unlocking the full potential of energy-time entanglement and pave the way for a resource-efficient, function-scalable, and highly compatible quantum network.

## Results

### 1. Principle and experimental setup

The energy-time entangled biphoton source (ET-EBS) has emerged as a versatile quantum resource, finding widespread applications across numerous quantum protocols. It offers the potential to concurrently implement multiple distinct quantum applications within a single quantum network. Here, we introduce a quantum network architecture as exemplified by Fig. 1(a), comprising two quantum nodes labeled as Alice and Bob. By leveraging multiplexed energy-time entangled biphotons and shared quantum link and infrastructures, this network enables the simultaneous realization protocols of two-way quantum time transfer (Q-TWTT) and unidirectional dispersive-optic quantum key distribution (DO-QKD). This achievement further provides robustness against potential asymmetric intercept-resend delay attacks from an eavesdropper, Eve.

As illustrated in Fig. 1(b), a proof-of-principle experiment is demonstrated between Alice and Bob nodes, departed by a 120 km single-mode fiber (SMF) link. The Q-TWTT configuration is established to obtain the clock offset and link delay information between the two nodes. This information is then utilized to synchronize their timing references, thereby ensuring the successful execution of the DO-QKD task. Two self-developed ET-EBSs[22] at 1560 nm were generated via the spontaneous parametric down-conversion (SPDC) process within the type-II MgO-doped periodically poled lithium niobate (MgO: PPLN) waveguide (refer to the "Methods" section for details). The idler photons (i1) originating from ET-EBS1 in Alice node are locally detected by the superconducting nanowire single-photon detector (SNSPD-D1, Photon Technology Co., Ltd.). Meanwhile, the accompanying signal photons (s1) that traverse the fiber link to Bob node are subsequently detected by SNSPD-D2. The arrival times of these photons, denoted as $t_{1,A}$ and $t_{1,B}$, are recorded by corresponding time tagger units (TTU1&TTU2, Time Tagger Ultra, Swabian Instruments), which are referenced to their individual nodes' clocks. Similarly, the emitted idler photons (i2) from ET-EBS2 at Bob node are locally detected by SNSPD-D3, while the signal photons (s2) travel through the same fiber link in the opposite direction and are then detected by SNSPD-D4 in Alice node. The arrival times of the photons from ET-EBS2 are denoted as $t_{2,A}$ and $t_{2,B}$, respectively. Through the equipped optical circulators (OC1 and OC2) in each node, the forward and backward transmitted signal photons (s1 and s2) share the same fiber link. To nonlocally compensate for the dispersion experienced by the signal photons traveling through the 120 km-long SMF, fiber Bragg grating-based dispersion compensation modules (DCM1&DCM2, DCMCB-SN-120P1FP, Proximion Inc.) were employed on the idler photon paths of both Alice and Bob nodes.

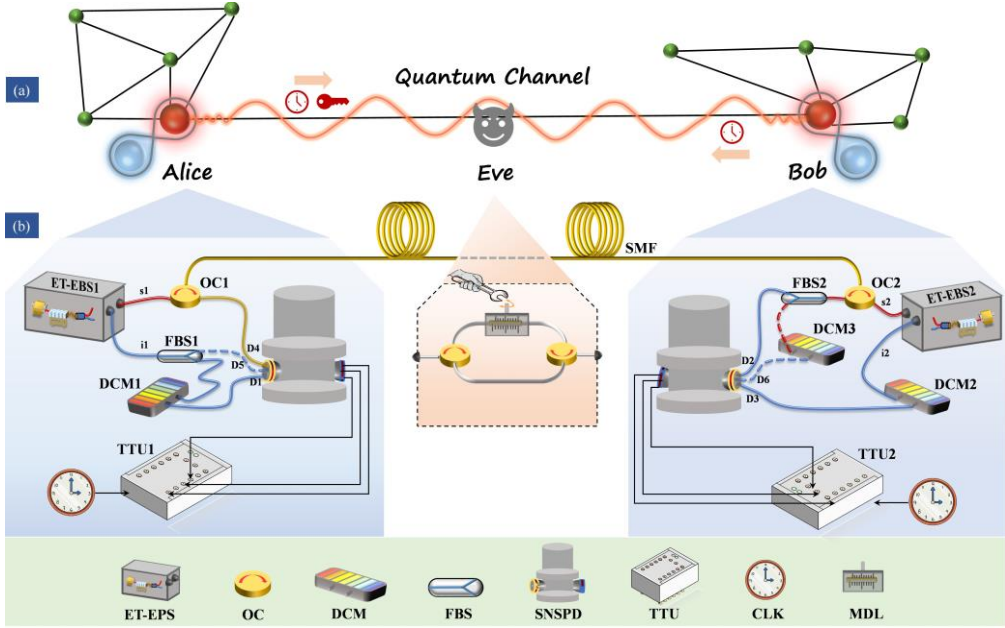

Fig. 1 Network architecture and experimental setup. (a) The network comprises two quantum nodes, Alice and Bob, enabling simultaneous bidirectional quantum time transfer and unidirectional quantum key distribution. The potential intercept-and-resend attacks posed by an eavesdropper, denoted as Eve. (b) A modified DO-QKD setup is integrated into the standard Q-TWTT framework. ET-EBS: custom-developed energy-time entangled biphoton source; FBS: fiber beam splitter; OC: optical circulator; DCM: dispersion compensation module; SNSPD: superconducting nanowire single-photon detector; TTU: time tagger unit; CLK: reference clock; MDL: motorized variable optical delay line.

Incorporating two fiber beam splitters (FBS1&FBS2) and a dispersion compensation module (DCM3) into the existing standard Q-TWTT framework, we have ingeniously adapted the setup to accommodate a modified version of DO-QKD. Note should be taken that, only one biphoton source (ET-EBS1) was employed for the DO-QKD process. In time bases, the single photon events are detected by SNSPD-D5 and SNSPD-D6, with local dispersion compensation being administered by DCM3. While via SNSPD-D1 and SNSPD-D2 the photon events are detected in frequency bases, where nonlocal dispersion cancellation is achieved with DCM1. Both the Q-TWTT and DO-QKD protocols concurrently utilize the arrival times of photons in frequency bases. Meanwhile, 70% of the photons in time bases are employed for key generation, and the remaining photons are used for security analysis. For conventional DO-QKD, the link dispersion is initially compensated, and then additional dispersion compensation modules with the opposite dispersion coefficient are used to perform separate transformations of the frequency basis for the two quantum nodes. The optimized scheme presented in this work significantly reduces the number of dispersion compensation components and streamlines the experimental setup.

To simulate the asymmetric delay attack from Eve, a setup comprising two optical circulators and one motorized variable optical delay line (MDL, MDL-002, General Photonics) was inserted into the fiber transmission link. By adjusting the relative time delay differences between forward and backward photons with MDL, quantifiable asymmetric time delay attacks were introduced in the bidirectional link.

## 2. Performance of Q-TWTT and DO-QKD

Firstly, the feasibility of the Q-TWTT protocol was verified by applying the 10 MHz signal of the Rb clock as the common frequency reference to Alice and Bob nodes. Taking the single run of measurement time as 5 seconds, the one-way link delay over 120 km fiber link was extracted by Gaussian fitting of the measured coincidence histograms, which are plotted in Fig. 2(a) with black line. Despite the two nodes using the same time reference, the link delay variation within half an hour reached 160 ps. By employing the Q-TWTT protocol, the symmetric delay in the bidirectional scheme was cancelled and the time offset can be deduced. The variation of $t_0$ as a function of measurement time is shown in Fig. 2(b) with red solid line, which follows a fluctuation as small as 1.9 ps in standard deviation. Correspondingly, Fig. 2(c) illustrates the time deviation (TDEV) results for the one-way link delay and time-offset data. As shown by black squares in Fig. 2(c), the link delay fluctuates significantly, especially in areas where the average time exceeds 50 s. For the time-offset, a TDEV of 1.7 ps was achieved at an averaging time of 5 s. It then exhibits a descending trend in accordance with the slope of $1/\sqrt{\tau}$, ultimately reaching 0.3 ps at 400 s.

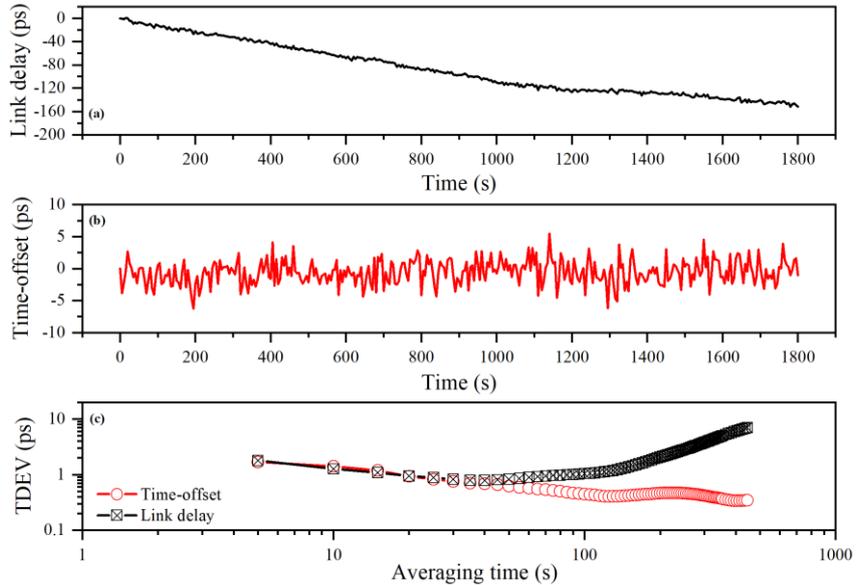

Fig. 2 Q-TWTT system performance with a common clock reference. (a) Measured one-way link delay. (b) Recovered time-offset by subtracting the bidirectional transmission delay. (c) The corresponding TDEV results.

For implementing the DO-QKD, the arrival times of the signal photons (s1) emitted from ET-EBS1 after transmission through the fiber link was corrected based on the extracted link delay and time-offset information. The coincidence distributions between the signal and idler photons before and after the correction are shown in Fig. 3(a) with green and orange bars, respectively. It is clear to see that, the tight temporal correlation of the distributed entangled photons is well preserved after 120 km-long fiber transmission to the distant node after the timing correction. A Gaussian fitting is implemented on the coincidence distribution, indicating a full width at half maximum (FWHM) of 83.6 ps, which is mainly attributed to the timing jitters of single photon detectors (FWHM ~76 ps) and a small amount of uncompensated dispersion of the optical fiber link. To obtain the noiseless system parameters used for calculating the excess noise factor and verify the security in the process of security test[18], the back-to-back experiment was configured by replacing the 120 km optical fiber link with another DCM (DCMCB-SP-120P1FP, Proximion Inc) of positive dispersion coefficient.

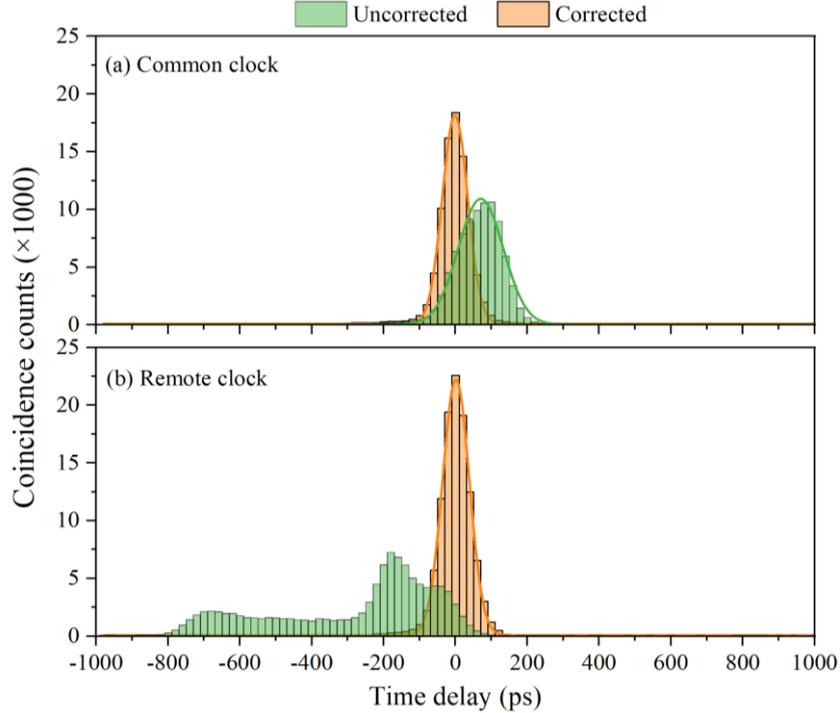

Fig. 3 Typical two-photon coincidence distributions in common clock (a) and remote clock (b) configurations. The orange (green) bars represent the coincidence distribution between Alice and Bob with (without) the link delay and time-offset information from the Q-TWTT. The Gaussian fitting curves are shown in the orange lines with an FWHM of 83.6 ps and 87.8 ps in (a) and (b), respectively.

A high-dimensional temporal encoding scheme is employed to efficiently increase the key rate by organizing the arrival time streams of photons into a structure of time frames, time slots and time bins.[23] To be specific, one time frame is segmented into $2^D$ time slots, and each time slot is further divided into $I$ time bins. Here, $D$ and $I$ represent positive integers that define the dimensions and the number of bins, respectively. The time bin, being the smallest unit of this temporal grid, has a width of $\tau$. Before the QKD process, Alice and Bob would determine their encoding parameters by a three-level optimization format.[23] When the encoding dimension or bin width is increased, the raw key rate (RKR) tends to rise. However, this improvement comes at the expense of a higher quantum bit error rate (QBER). Consequently, it is essential to strike a balance between the RKR and QBER. In the current experiment, the QBER is set to below an upper bound of 5% along with optimized encoding parameters of $D=6$, $I=3$ and $\tau = 110$ ps. The obtained RKR is 199.4 bits per second (bps) with a QBER of 4.9%. When considering the finite-size effect with the total coincidence counts of about $3\times10^5$, the secure key capacity is 3.52 bits per coincidence (bpc) and the secure key rate (SKR) is 117.0 bps.

To shift from the common-clock scenario to a more practical situation in which a remote clock synchronization between Alice and Bob is accomplished, we adopted a scheme involving open-loop fiber-optic microwave frequency transfer and dynamic phase compensation.[24] Despite observing some degradation in the coincidence distribution, the achieved clock synchronization enabled us to nearly revert to the coincidence distribution observed with a common clock reference. As shown in Fig. 3(b), a FWHM of 87.8 ps was achieved. In this case, DO-QKD was performed with re-optimized parameters of $D=6$, $I=3$ and $\tau=80$ ps. The obtained RKR is 181.6 bps with a QBER of 4.8% and the finite-size SKR is determined to be 100.5 bps.

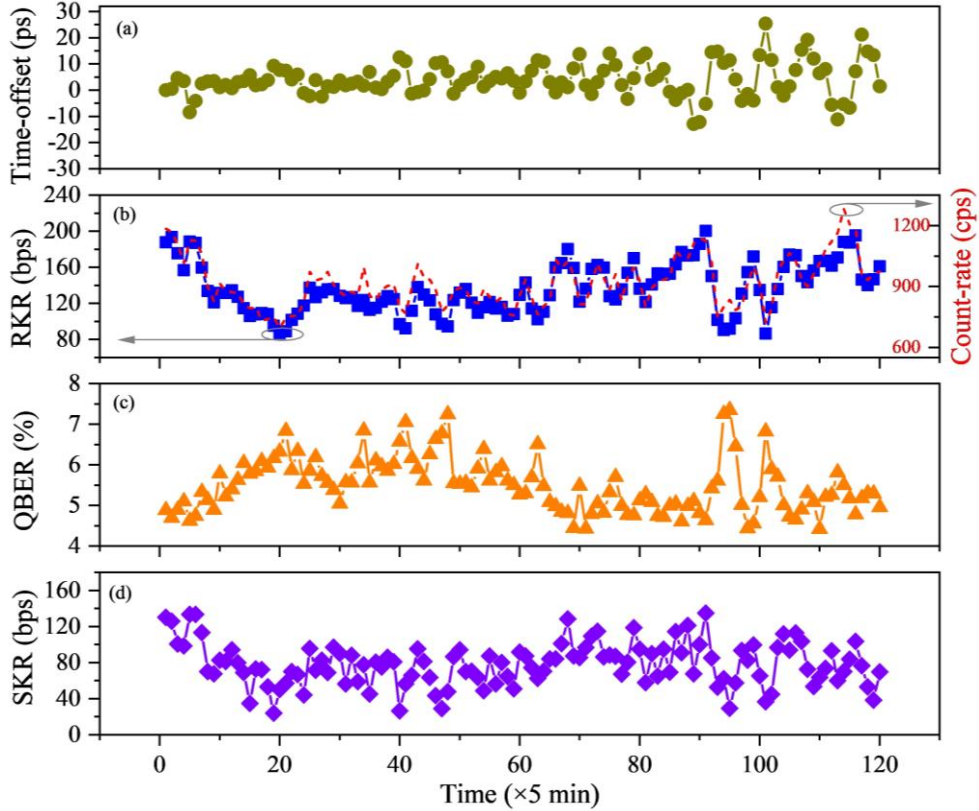

Fig. 4 Time traces of the time-offset (a), RKR along with the single-side count-rate (b), QBER (c) and asymptotic SKR (d). Each point represents the average value over 5 minutes of raw data.

In the remote clock synchronization scenario, the long-term performance analysis of Q-TWTT and DO-QKD lasting for 10-hour was further conducted. The collected raw timestamp data was segmented into 5-minute intervals for detailed analysis, and time traces of the time-offset, RKR, QBER and asymptotic SKR are shown in Fig. 4 (a)-(d), respectively. As shown in Fig. 4(a), the time-offset between the two nodes Alice and Bob is stabilized with a residual fluctuation of 6.5 ps in standard deviation. Leveraging the time synchronization information, the RKR was sustained at an average of (136.7±27.9) bps as seen in Fig. 4(b). Note should be taken that, in the initial data segment (spanning 0-100 minutes), the time-offset remains very stable, which suggests excellent time synchronization performance between the two nodes. However, the RKR shows apparent drop from ~200 bps to ~80 bps. As depicted by the red dashed line in Fig. 4(b), the RKR exhibits the same trend as the photon count rate. This observation suggests that the fluctuation in photon counts serves as the primary factor contributing to the variation in the key rate. The underlying reason is that long-haul optical fiber transmission causes random polarization changes in photons, and the utilized SNSPDs are polarization-sensitive. On the other hand, since the accidental coincidence count rate remains largely stable throughout the testing process, a lower RKR results in a higher QBER, as shown in Fig. 4(c). Consequently, even though the QBER was optimized to fall below the predetermined threshold of 5% before initiating the long-term measurement, the average QBER ultimately registers at (5.5±0.7)% due to the inherent fluctuations. The asymptotic SKR is then calculated, as shown in Fig. 4(d), indicating a similar trend as the RKR. Furthermore, the finite-size SKR was determined to be (73.8±15.7) bps throughout the long-term test. Remarkably, even when simultaneously operating dual quantum protocols via multiplexed energy-time entangled biphotons, the achieved SKR demonstrates performance competitiveness against state-of-the-art standalone entanglement-based QKD systems over fiber links.

## 3. Robustness against asymmetric delay attacks

As illustrated in Fig. 3, precise time synchronization is essential for reestablishing the temporal correlation of two-photon entanglement, which forms the cornerstone of DO-QKD. In practical applications, the presence of an eavesdropper who introduces asymmetric time delay attacks into the transmission link can lead to erroneous timing information, consequently degrading the performance of DO-QKD. To assess the impact of asymmetric delay attacks on the DO-QKD, we configured Alice and Bob nodes with a common clock reference to emulate a scenario without the quantum time synchronization (QTS) procedure. The time delay attack manipulated by Eve was simulated by incrementally setting the MDL-induced delay from 0 ps to 120 ps, with a step of 10 ps. At each delay setting, the system was run for 5 minutes. As shown in Fig. 5(a), the statistical histogram of the measured time-offset features a flat-top shape, suggesting that Eve has introduced a substantial timing shift. Subsequently, the relationship between the measured normalized SKR and the quantity of the introduced time delay attacks is obtained and depicted in Fig. 5(b) with purple diamonds. The normalized SKR exhibits a progressive decline with increasing asymmetric time delay, demonstrating a pronounced reduction to 13% when the asymmetric delay reaches 120 ps. To explain this phenomenon, we introduced corresponding varying time delays to the experimentally measured raw timing data from the baseline configuration (i.e., without an asymmetric delay attack). This methodology enables quantitative simulation of how asymmetric delay attacks compromise the key generation efficiency in DO-QKD. The simulation results, depicted by red squares in Fig. 5(b), reveal the dependence of the normalized SKR on the asymmetric time delay. The experimentally measured data points show strong agreement with the simulation results, with residual discrepancies primarily arising from the inherent photon count rate fluctuations discussed in the previous section. These findings demonstrate that asymmetric time delay attacks on the time synchronization system are highly detrimental to the performance of DO-QKD.

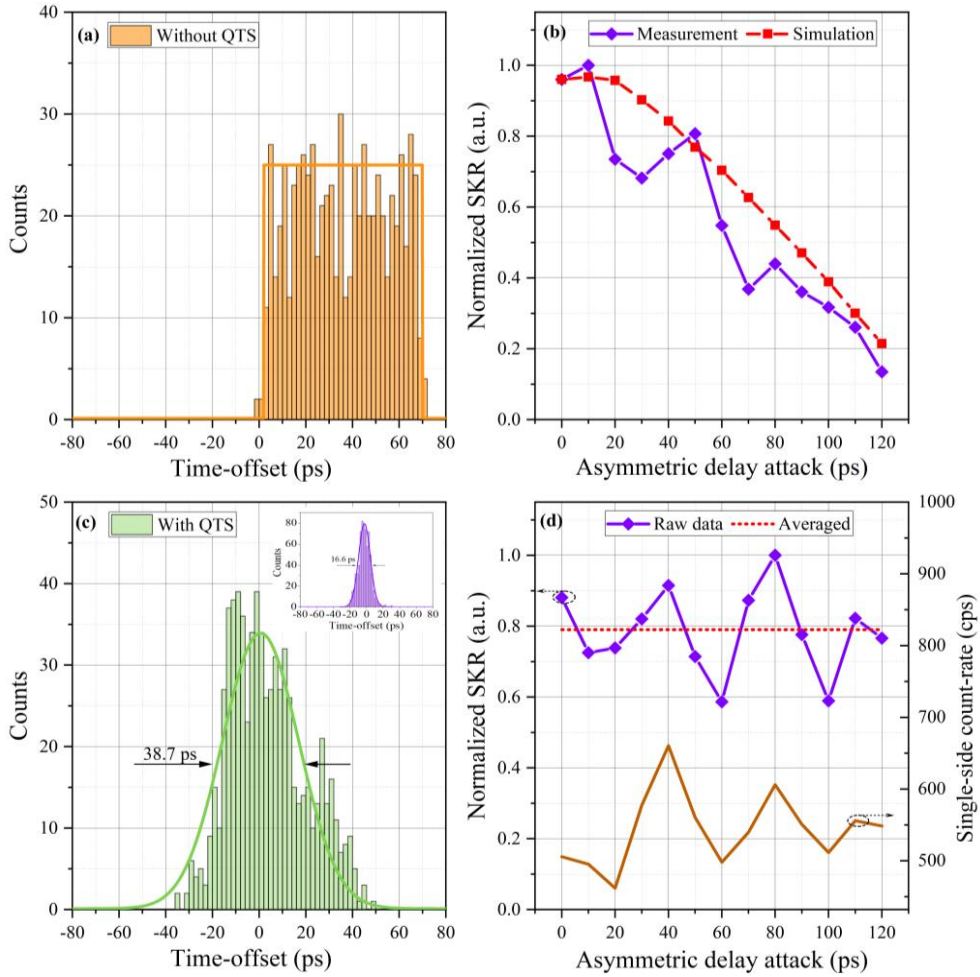

Fig. 5 Statistical histograms of the time-offset and normalized SKR in asymmetric delay attack scenarios, with and without QTS. (a) Without QTS: Statistical histograms of the measured time-offset data, with the rectangular fitting envelope shown as a solid line. (b) Without QTS: The graph of the normalized SKR as a function of the quantity of introduced asymmetric delay attacks, with the measured data plotted by purple diamonds and simulated data represented by red squares. (c) With QTS: statistical histograms of the measured time-offset data, with the Gaussian fitting envelope shown by the solid line; the insert shows the histogram under the no-attack condition. (d) With QTS: The graph of the normalized SKR as a function of the quantity of introduced asymmetric delay attacks, with the measured data plotted by purple diamonds, the average value indicated by the red dashed line, and the detected single photon count-rate at Bob node shown by the solid crayon-colored line.

To mitigate these risks and ensure the continued efficacy of DO-QKD, it is imperative to enhance the robustness of the system against asymmetric delay attacks. According to the protocol described in this paper, QTS and DO-QKD make use of the same energy-time entangled photons and pass through the identical fiber link. Thus, asymmetric delay attacks can be regarded as a form of common-mode noise for the two applications, which is expected to have no impact on the performance of DO-QKD. To elaborate on this point, let us first consider the scenario in the absence of attack. Suppose $t_A$ and $t_B$ represent the reference times at the Alice and Bob nodes respectively, $\tau_{\text{link}}$ denotes the symmetry link transmission delay. Since the QTS process aims to synchronize the time reference at the Bob node with that at the Alice node, the true time-offset extracted based on the Q-TWTT protocol, namely $t_0 = t_B - t_A$, is the value that needs to be subtracted to $t_B$ to achieve synchronization. In other words, the new time reference at Bob ($t'_B$) should be adjusted such that $t'_B = t_B - t_0 = t_A$. Next, we consider the situation where an asymmetric delay attack, denoted as $\tau_{Eve}$, is introduced by Eve in the forward path from Alice to Bob. According to the working principle of the Q-TWTT protocol (see Methods for details), a false time-offset ($t'_0$) is extracted,

$$t'_0 = (t'_B - t_A) + \frac{\tau_{Eve}}{2}. \tag{1}$$

Given that the QTS process continuously synchronizes the reference clock of the Bob node with that of the Alice node, making $t'_0$ approach 0, the new time reference at Bob node is then readjusted to:

$$t''_B = t'_B - t'_0 = t_A - \frac{\tau_{Eve}}{2}. \tag{2}$$

Simultaneously, the measured link delay ($\tau''_{\text{link}}$) is determined as

$$\tau''_{\text{link}} = \tau_{\text{link}} + \frac{\tau_{Eve}}{2}. \tag{3}$$

From Eqns. (2) and (3), it can be observed that the asymmetric delay attack has an equal but opposite effect on the time reference established via QTS ($t''_B$) and the measured link delay ($\tau''_{\text{link}}$). For the implementation of the DO-QKD, the arrival time of the signal photons (s1) at Bob is determined by $t''_B + \tau''_{\text{link}} = t_A + \tau_{\text{link}}$, which aligns with the scenario where no asymmetric delay attack is present.

To validate this conclusion, we adopted the aforementioned scheme to conduct corresponding tests. Even in the presence of asymmetric delay attacks, the internal time offset of the system invariably converges towards a fixed position, as illustrated by the histogram in Fig. 5(c). Despite some broadening relative to the no-attack condition (insert), the histogram retains a distinct Gaussian envelope, with a fitted FWHM of 38.7 ps. The performance of DO-QKD was further analyzed, by incrementally varying the MDL-induced delay from 0 ps to 120 ps, the resultant finite-size SKR in its normalized form is depicted in Fig. 5(d) with purple diamonds. In contrast to Fig.5 (b), with the increase of the asymmetric delay attack, no decreasing trend can be observed. Instead, the values fluctuate around an average of 0.79. This fluctuation can be attributed to the fluctuation of the single-side photon count-rate at Bob's node. As shown by the solid crayon-colored line in Fig. 5(d), it exhibits a consistent variation pattern with that of the SKR. The results unambiguously demonstrate that detrimental effect on the DO-QKD induced by the asymmetric delay attacks in the fiber link can be effectively mitigated through the concurrent QTS process. This finding underscores the crucial role of entanglement multiplexing in enhancing the robustness against asymmetric time delay attacks, thereby ensuring the reliable and secure operation of quantum networks.

## Discussion

By multiplexing the energy-time entanglement, two distinct quantum protocols, quantum time synchronization and dispersive-optic quantum key distribution, are concurrently enabled. A proof-of-principle demonstration between two independent nodes across a 120 km fiber-optic link was implemented. Through the Q-TWTT protocol, we achieved time synchronization reaching a stability well below 1 ps. It then provides the critical prerequisite for the execution of DO-QKD, which succeeded in generating a finite-size secret key rate of (73.8±15.7) bps. During each 5-second interval dedicated to implementing a single-rum of QCS procedure, a secure key comprising approximately 369 bits is generated. This key is adequate to support the widely-adopted Advanced Encryption Standard 256 (AES-256), thereby enhancing the data layer security of the QCS system. By obviating the necessity for transmitting classical time synchronization signals, our protocol surmounts the challenge of co-fiber transmission of classical and quantum signals. Furthermore, benefiting from the use of energy-time entangled photon pairs as the common quantum resource and the shared quantum infrastructures, the presented DO-QKD showcases robustness against potential asymmetric time delay attacks in the transmission link: the normalized SKR has been verified to remain fairly constant around 80% even under a malicious delay attack up to 120 ps.

One may notice that the generated secret key rate based on the DO-QKD protocol experiences declining sometimes despite stable synchronization between the two nodes. This is found to be mainly associated with fluctuations in the detected photon counts, which should be attributed to the contradiction between the fiber-induced polarization fluctuations of the transmitted signal photons and the high polarization sensitivity of the SNSPDs. To address this issue, advanced polarization management and photon detection techniques should be adopted, including the use of electrical polarization controllers to mitigate polarization fluctuations and the creation of polarization-insensitive SNSPDs, both of which would substantially improve the QKD performance. These achievements represent a significant milestone in the utilization of energy-time entanglement towards establishing a foundation for a resource-efficient, function-scalable, and highly compatible quantum network. Benefitting from the versatile capability of the energy-time entanglement photon pair source, new possibilities for integrating more quantum information applications in a single quantum network can be highly expected.

## Methods

### Energy-time entangled biphoton source

The all-fiber energy-time entangled biphoton sources (ET-EBS) were self-developed based on the spontaneous parametric down-conversion (SPDC) process.[22] Paired signal and idler photons were produced from a type-II MgO-doped periodically poled lithium niobate (MgO: PPLN) waveguide, pumped by a single-mode fiber pigtailed laser diode (Thorlabs, DBR-780PN) at 780 nm. Afterwards, a customized filtering module was connected behind the output of the PPLN waveguide to eliminate the residual 780 nm pump beam. The signal and idler photons were then spatially separated by a fiber polarization beam splitter. To avoid the polarization-dependent instability, all the above-mentioned fiber components were polarization-maintaining. Despite the center wavelength of the two ET-EBSs is set to be around 1560 nm, the difference in PPLN length results inconsistent spectral widths of 1 nm and 1.7 nm in FWHM.

### Quantum two-way time transfer protocol

The scheme for realizing the quantum two-way time transfer[13] between two clocks at separate nodes (Alice and Bob) that are interconnected via a fiber link. Each node has an energy-time entangled biphoton source, single-photon detectors, and a time tagger unit referenced to its local timescale. For the entangled biphoton source generated at Alice node, the signal photons travel from Alice to Bob through a fiber link with transmission delay of $\tau_{link,AB}$, while the idler photons are detected locally at Alice node. Using the time tagger units, the times of arrival for the detected signal and idler photons are then recorded as $\{t_{1,B}^{(j)}\}$ and $\{t_{1,A}^{(j)}\}$ respectively, where $j = 1, 2, 3 \dots n$ denotes a series of time tagged sequences. By applying a cross-correlation algorithm, the coincidence histogram of the time difference between $t_{1,B}^{(j)}$ and $t_{1,A}^{(j)}$ can be constructed. Through Gaussian fitting of the coincidence distribution, the registration time difference $t_1 = t_{1,B}^{(j)} - t_{1,A}^{(j)}$ with respect to the maximum coincidences is obtained. Assume the time difference between clocks at Bob and Alice nodes is $t_0$, it can be deduced that $t_1 = \tau_{link,AB} + t_0$. For the energy-time entangled biphoton source at Bob node, a similar procedure is carried out with recorded time stamps of $\{t_{2,A}^{(j)}\}$ and $\{t_{2,B}^{(j)}\}$ and the time difference ($t_2 = t_{2,A}^{(j)} - t_{2,B}^{(j)}$) can be extracted $t_2 = \tau_{link,BA} - t_0$. Under the condition of link symmetry ($\tau_{link,AB} =$ $\tau_{link,BA} = \tau_{link}$), the time-offset between the two clocks is then given by $t_0 = (t_1 - t_2)/2$. At the same time, the link delay can be also obtained with $\tau_{link} = (t_1 + t_2)/2$. In the remote time synchronization process[24], the reference clock at Bob node is fine-tuned according to the measured time-offset, which is executed by a programmable phase trimmer.

**DO-QKD protocol**

In the DO-QKD protocol, we employ a high-dimensional encoding format[23] to generate keys more efficiently. Based on the acquired timing information from the Q-TWTT protocol, the arrival time of signal photons after fiber transmission are corrected. The calibrated arrival times of photons are divided into time frames, time slots, and time bins. A time frame has $2^D$ time slots, a time slot has $I$ time bins, and a time bin has a width of $\tau$. Before the QKD process, Alice and Bob would determine their encoding parameters by a three-level optimization format. In the bin-sifting process, each user first selects the frames with only one single-photon event and discards other frames. Then they exchange the frame number and bin number of each preserved single-photon event through classical communication. Only photons from two users have the same frame number and the same bin number would be paired and used to generate keys. Then the raw keys are created using the slot numbers of the single-photon events.

The security analysis of DO-QKD relies on the treatment of single-photon events and the well-established proofs of Gaussian CV-QKD.[18] Through the calculation of the time-frequency covariance matrix (TFCM) between photons' arrival times of Alice and Bob, the security of protocol against Eve's Gaussian collective attacks has been proven. The secure key capacity represents the number of secure keys that can be extracted from each coincidence count, which is denoted as:

$$\Delta I = \beta I(A;B) - \chi(A;E) - \Delta_{FK}, \quad (4)$$

where $\beta$ is the reconciliation efficiency, $I(A;B)$ is Shannon mutual information between Alice and Bob, $\chi(A;E)$ is Eve's Holevo information[18], and $\Delta_{FK}$ accounts for the penalty of the finite-size effect. Considering the finite length of generated keys, each stage of the QKD protocol has a probability of failing. The tolerated failure probability of the whole protocol $\varepsilon_s$ is denoted as $\varepsilon_s = \varepsilon_{ver} + \varepsilon_{PA} + n_{PE}\varepsilon_{PE} + \bar{\varepsilon}$. It is the sum of failure probabilities in stages of error correction and verification, privacy amplification, parameter estimation, and estimating the smooth min-entropy. In the test, the reconciliation efficiency is taken as 90% based on our previous experimental results.[25]

## Data availability

The raw data that support the findings of this study are available from the corresponding author upon reasonable request.

## Code availability

The codes that have been used for this study are available from the corresponding author upon reasonable request.

## Acknowledgements

We thank Prof. Gerd Leuchs for enlightening discussions. This work is supported by National Natural Science Foundation of China (12033007, 61801458, 12103058, 12203058, 12074309, 61875205, 92365210), Youth Innovation Promotion Association of the Chinese Academy of Sciences (2021408, 2022413, 2023425), and the Innovation Program for Quantum Science and Technology (2021ZD0300900).

## Author contributions

R. D. and W. Z. conceived the original idea and supervised the work; X. X. and J. L. designed the experiment, collected data and analyzed the data; B. S., H. H. and X. S. participated in part of the experiment for data collection; Y. L. built the energy-time entangled biphoton source; X. X. and J. L. prepared the original manuscript. R. D., W. Z., X. X., J. L., R. Q., T. L., and S. Z. revised the manuscript.

## Competing interests

The authors declare no competing interests.